# Continuous SpO$_2$ Monitoring Using Reflectance Pulse Oximetry at the Wrist and Upper Arm During Overnight Sleep Apnea Recordings

Karen Adam[*,1], Clémentine Aguet[1], Patrick Theurillat[1], Florent Baty[2], Maximilian Boesch[2], Damien Ferrario[1], Mathieu Lemay[1], Martin Brutsche[2], Fabian Braun[1]

*Abstract*— Sleep apnea (SA) is a chronic sleep-related disorder consisting of repetitive pauses or restrictions in airflow during sleep and is known to be a risk factor for cerebro- and cardiovascular disease. It is generally diagnosed using polysomnography (PSG) recorded overnight in an in-lab setting at the hospital. This includes the measurement of blood oxygen saturation (SpO$_2$), which exhibits fluctuations caused by SA events. In this paper, we investigate the accuracy and utility of reflectance pulse oximetry from a wearable device as a means to continuously monitor SpO$_2$ during sleep. To this end, we analyzed data from a cohort of 134 patients with suspected SA undergoing overnight PSG and wearing the watch-like device at two measurement locations (upper arm and wrist). Our data show that standard requirements for pulse oximetry measurements are met at both measurement locations, with an accuracy (root mean squared error) of 1.9% at the upper arm and 3.2% at the wrist. With a rejection rate of 3.1%, the upper arm yielded better results in terms of data quality when compared to the wrist location which had 30.4% of data rejected.

*Clinical Relevance*—Our findings confirm the feasibility of wearables for unobtrusive continuous monitoring of SpO$_2$, especially in patients with suspected SA.

## I. INTRODUCTION

Sleep apnea (SA) is a chronic sleep-related disorder estimated to affect 23% of the general population worldwide and is a risk factor for developing cerebro- and cardiovascular comorbidities. SA consists of repetitive pauses or restrictions in breathing during sleep, which in turn influence the levels of blood oxygen saturation. One way to diagnose SA is to track fluctuations in peripheral blood oxygen saturation (SpO$_2$) during sleep, which can be measured using fingertip photoplethysmography (PPG) acquired during a polysomnography (PSG).

While this diagnostic modality is generally performed overnight at the hospital, we seek to offer a more scalable solution applicable to ambulatory care, by means of a watch-like device wearable at the wrist or upper arm. Such an approach could reduce costs as compared to standard in-hospital PSG.

To investigate this, we explored data obtained from a cohort of 134 patients with suspected SA who have undergone PSG. Using a fingertip pulse oximeter device as reference, we evaluated the accuracy of a proprietary wearable device continuously providing SpO$_2$ estimates, both at the upper arm and wrist, using 3 second and 15 second windows (Figure 1). We examine performance results for SpO$_2$ estimation at both the wrist and upper arm and show (1) preliminary compliance with major pulse oximetry standards and (2) a discrepancy in performance between SpO$_2$ estimated at the upper arm vs wrist.

[*] Corresponding author: karen.adam@csem.ch.
[1] Centre Suisse d'Electronique et de Microtechnique (CSEM), Neuchâtel, Switzerland
[2] HOCH Health Ostschweiz, Kantonsspital St.Gallen, Universitäres Lehr- und Forschungsspital, Lung Center, St.Gallen, Switzerland.

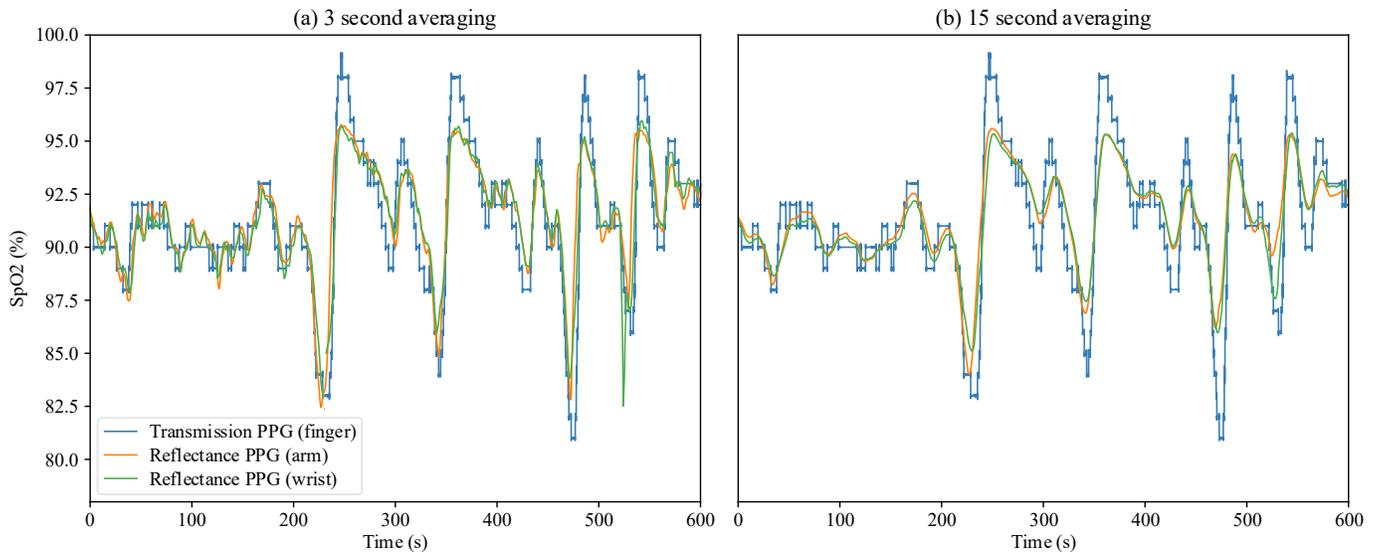

Figure 1. Comparison of SpO$_2$ estimate obtained from (blue) reference fingertip PPG device, (orange) CSEM reflectance PPG wearable at upper arm and (green) CSEM reflectance PPG device at wrist, using averaging windows of 3 seconds (left) and 15 seconds (right). The comparison between the two averaging windows shows that a 3-second averaging windows allows better tracking of the dynamics of SpO$_2$ levels, and the comparison between the arm and wrist PPG estimate show that they can both provide reasonably good estimates of SpO$_2$ provided sufficient PPG data quality.

## II. METHODS

### A. Study

Subjects with a suspicion of SA underwent overnight PSG as part of the routine diagnostic process. On this occasion, reflectance PPG signals were additionally collected at the upper arm and wrist using a proprietary wearable device. The data in question refer to a subset of a wider data collection campaign currently taking place at the HOCH Health Ostschweiz which encompasses 700 subjects, and which was approved by the local ethics committee (Ethikkommission Ostschweiz, BASEC-ID: 2024-00637).

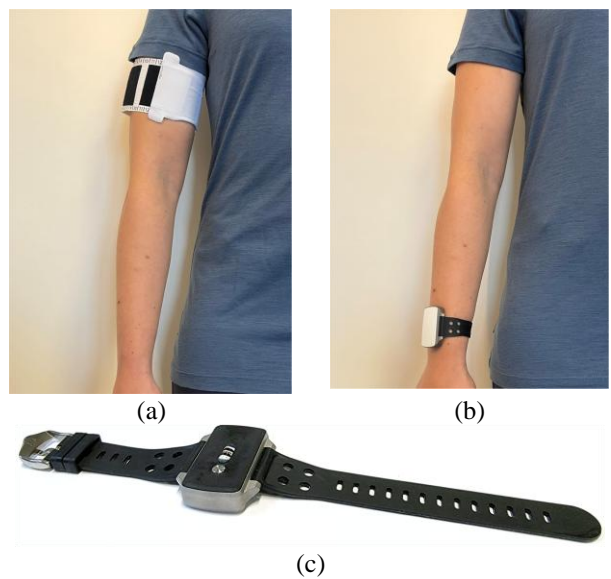

Figure 2: CSEM wearable worn at (a) upper arm and (b) wrist and (c) back view (skin-facing side) of wearable.

At the time of data lock, data from 134 of these 700 patients were available for analysis with further statistics about the cohort listed in TABLE I.

TABLE I. COHORT STATISTICS (N=134)

| Characteristics | Median (IQR) or count (%) |
|---|---|
| Age (y) | 49 (38.3 - 60) |
| Gender, male (count) | 99 (73.9) |
| Body Mass Index (kg/m$^2$) | 29.5 (26.3 - 33.6) |
| Skin Type, dark (count)[3] | 27 (20.1) |
| SpO$_2$ Reference Value (%) | 93.0 (92.0 – 95.0) |
| Apnea-Hypopnea Index (1/h) | 30 (12 - 55.3) |
| Epworth Sleepiness Scale (-) | 8 (5 - 13) |

### B. Measurement and Reference Devices

The wearables used to measure reflectance PPG are proprietary devices from CSEM (Neuchâtel, Switzerland) and are shown in Figure 2c. One of them was placed at the wrist using a wristband (Figure 2b), while the other was placed at the upper arm using an armband (Figure 2a).

Each of these devices emits light in the red and infrared ranges (660 nm and 950 nm respectively), at a frequency of 50 Hz, as well as light in the green range (525 nm), at a frequency of 100 Hz. While the SpO$_2$ algorithm can be embedded into the device, in this study, the PPG data were stored locally on-device and later exported for offline processing and performance evaluation. This was done to ensure signal alignment between the wearable and reference device, and to evaluate SpO$_2$ estimation performance under different algorithm settings.

The SpO$_2$ reference device used was a fingertip PPG device (Xpod 3012LP, Nonin Inc., Plymouth, USA) which outputs SpO$_2$ values at a rate of 1 Hz (with a maximal averaging time of 3 seconds). We also used the electrocardiogram (ECG) signal simultaneously acquired with this reference SpO$_2$ to time-synchronize the PSG data with the two wearable devices, based on inter-beat intervals.

### C. Applicable Standards

Two categories of medical device standards apply to this SpO$_2$ evaluation study.

The first category relates to pulse oximetry in general and includes the ISO standard [3] as well as FDA guidance [4] for reflectance pulse oximetry. With regards to accuracy, they both encourage the use of accuracy root mean squared (A$_{RMS}$) error as a metric of accuracy. The A$_{RMS}$ should take blood samples as a reference, in which blood samples are drawn during stable SpO$_2$ plateaus that last longer than 30 seconds, and that follow a hypoxia protocol. The ISO standard (resp. FDA) specifies a maximal A$_{RMS}$ error of 4.0% (resp. 3.5%) under such conditions for a SpO$_2$ range of [70%-100%]. In addition, both standards dictate using data no older than 30 seconds for every value of SpO$_2$ output by the pulse oximeter in question.

---

[3] For the first 32 patients, skin tone was assessed manually using Fitzpatrick or Monk Skin Tone Scale. For the remaining 102 subjects, skin tone was quantified as individual typology angle (ITA) using a colorimeter (SkinPhotoTyper, Delfin Technologies, Kuopio, Finland). To ensure consistency in the analysis, the different skin type measurements were converted into a unified Boolean metric (ITA $< 10°$, Fitzpatrick Scale $> 5$, and Monk Skin Tone Scale $> 7$), indicating whether the skin type is classified as dark [1], [2].

On the other hand, the American Association of Sleep Medicine (AASM) specifies requirements on the various vital sign monitoring devices used during PSG [5]. When it comes to pulse oximeters, it requires an averaging time of no more than 3 seconds, unless the pulse rate is lower than 80 beats per minute, in which case the $SpO_2$ value can be averaged over the duration of three pulses.

## III. $SpO_2$ ESTIMATION AND EVALUATION

### A. $SpO_2$ Estimation

For each of the wearable devices, $SpO_2$ estimation was performed by taking the red, infrared and green PPG signals resampled at 50 Hz and obtaining a ratio of ratios value (ROS) for the red and infrared channels on a pulse-by-pulse basis, which is then averaged over each window. The estimated $SpO_2$ value is obtained through an affine transformation applied to the ROS value (where lower ROS values result in higher $SpO_2$ values). This transformation is device-specific and was calibrated to the wearable device using data previously obtained at both the upper arm and wrist during a hypoxia protocol, as specified by the validation protocol of the ISO 80601-2-61:2017 standard [3].

Moreover, the $SpO_2$ estimation algorithm returns a quality index (QI) which indicates signal quality and confidence in the $SpO_2$ estimation. We applied an empirically derived threshold to this quality index to exclude potential erroneous $SpO_2$ estimates from evaluation.

We evaluated the algorithm performance under two different settings of averaging window duration: a) 15 seconds as well as b) 3 seconds (to be compliant with the AASM guidelines [5]), both with a 1-second refresh rate (i.e., the sampling rate of $SpO_2$ estimates). While a) is expected to lead to estimates with lower overall error, b) is likely to better follow fast changes due to apneic events.

As previously mentioned, although $SpO_2$ estimation can be embedded in the device for a real-time output, we performed the analysis offline to (1) align estimates and reference signals for a fair evaluation and (2) perform the analysis with different algorithm settings.

To account for delays caused by differences in processing or averaging times between $SpO_2$ reference (fingertip, calculated online) and estimates (wrist or upper arm, calculated offline), we further accounted for a delay in the $SpO_2$ value output by the fingertip PPG device. To this end, we empirically determined the delay between fingertip and wearable $SpO_2$ for each measurement location (upper arm and wrist) separately, by minimizing the $A_{RMS}$ error with respect to the time delay.

### B. Statistical Analysis

To evaluate the performance of our $SpO_2$ estimation algorithm, we considered two main metrics, the accuracy as root mean squared error ($A_{RMS}$), and the acceptance rate.

The $A_{RMS}$ error was computed by comparing the fingertip $SpO_2$ value with our output value, for sections of the signal where both the reference and the estimate were deemed to be of sufficient quality. The reference was deemed to be of sufficient quality if it varied by less than 3% over windows of 5 seconds, and if the values were in the accepted range of 60% to 100% $SpO_2$. The estimate was deemed to be of sufficient quality if it was in the same accepted range, and the quality index output by our algorithm was lower than an empirically set threshold. This quality index assessed the likelihood of each PPG pulse being of physiological origin, as well as the level of noise recorded.

As for the acceptance rate, we report the proportion of the signal where the estimate was not rejected due to too low QI despite the reference value being of sufficient quality. Note that the reference rejection accounts for the rejection of 6% of all signals regardless of QI settings, and it was not accounted for in the acceptance rate listed in the performance tables included later on.

All reported values were computed by concatenating the portions of each recording which were not rejected, so that the average was performed over all non-rejected samples, as opposed to being done over all subject $A_{RMS}$ error values.

Agreement between the reference and estimated $SpO_2$ values was further examined using Bland-Altman analysis.

## IV. RESULTS

TABLE II. shows the overall performance metrics for both 15- and 3-second averaging durations.

All results were obtained using time delays of 9 seconds for the upper arm and 5 seconds for the wrist compared to the fingertip $SpO_2$ value.

Figure 3 shows Bland-Altman analyses comparing $SpO_2$ estimates with the $SpO_2$ reference, across different $SpO_2$ levels.

We also provide an example in Figure 1, where we plot $SpO_2$ estimates at each location, and computed using 3 second and 15 second averaging, against the reference $SpO_2$, for one subject undergoing apneic events.

TABLE II.  PERFORMANCE OF $SpO_2$ ESTIMATION

| **Performance Metric** | **Averaging Window Duration** | | | |
|---|---|---|---|---|
| | *a) 15 seconds* | | *b) 3 seconds* | |
| | Wrist | Upper Arm | Wrist | Upper Arm |
| $SpO_2$ $A_{RMS}$ Error | 3.2% * | 1.9% * | 3.5% * | 2.0% * |
| Error Bias | -0.7% | 0.4% | -0.8% | 0.4% |
| Acceptance Rate | 69.6% | 96.9% | 68.3% | 96.0% |

\* Compliant with ISO standard [6] and FDA guidance [4] for reflectance type sensors.

## V. DISCUSSION

In this study, we evaluated the performance of our $SpO_2$ estimation based on reflectance PPG acquired using wearables at the upper arm and wrist in 134 SA patients during overnight PSG.

### A. Performance of Reflectance $SpO_2$ with Wearables

For both measurement locations and averaging durations studied, the $A_{RMS}$ error obtained after evaluation of our $SpO_2$ estimate (TABLE II. ) was lower than the maximal $A_{RMS}$ error permitted by both the ISO standard [6] ($\leq 4.0\%$) and the FDA guidance for reflectance pulse oximetry [4] ($\leq 3.5\%$). However, for the two averaging durations investigated, we observe an increase in both $A_{RMS}$ error and rejection rate when measuring reflectance $SpO_2$ at the wrist when compared to the upper arm.

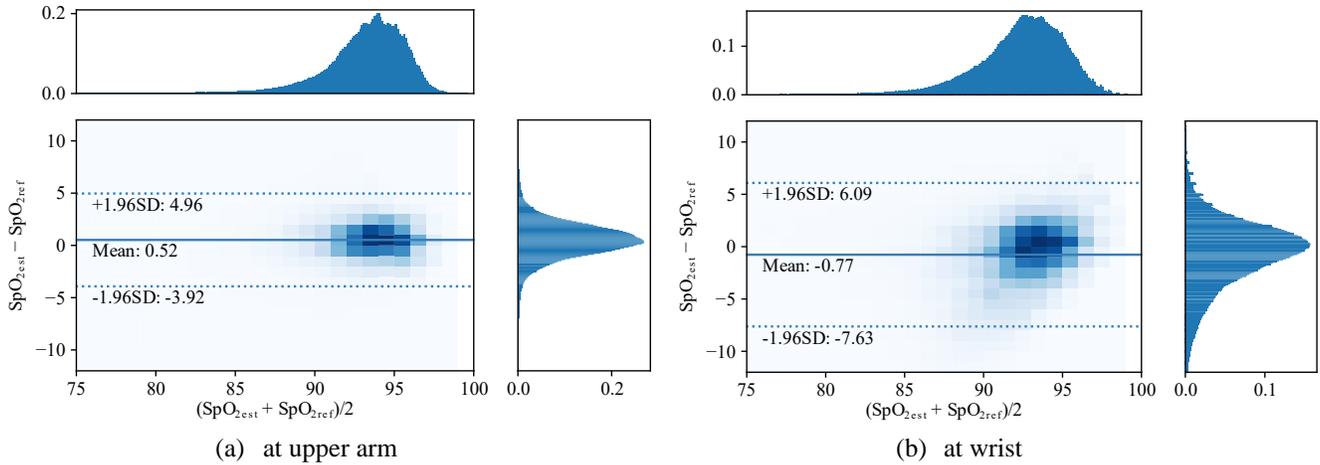

Figure 3: Bland-Altman plots displaying error based on $SpO_2$ value when estimating $SpO_2$ using CSEM proprietary reflectance PPG device at (a) upper arm and (b) wrist with 3 second averaging windows

These discrepancies in performance between reflectance $SpO_2$ estimation at the upper arm versus the wrist – with performance at the wrist being noticeably worse – is in line with observations from a previous study in a small group of nine sleep apnea patients [7].

Moreover, $SpO_2$ estimate bias values obtained at the upper arm and wrist differ by approximately 1.2%, with estimates at the wrist and upper arm exhibiting negative and positive bias, respectively. This is also in line with previous observations [7]. We hypothesize that at the wrist, increased levels of noise tend to bring ROS values closer to 1 and thus tend to lead to underestimated $SpO_2$ values.

## B. Issues of Measurements at the Wrist

We believe the discrepancy of results is due to a confluence of factors.

First, the placement of a wearable at the wrist is more likely to be adversely impacted by the surface bone structure at that location, which results in a less tight fit, and thus a potentially higher level of noise, movement artifacts or a deteriorating influence of venous blood or tendons.

Second, a wearable placed at the wrist is more often impacted by movement, as a person tends to move their forearm more than their upper arm, which in turn influences the PPG signal quality.

These factors result in overall lower SNR at the wrist which leads to an increase in the $A_{RMS}$ error, but which is also reflected in a decrease in quality index and an increase in resulting rejection rate.

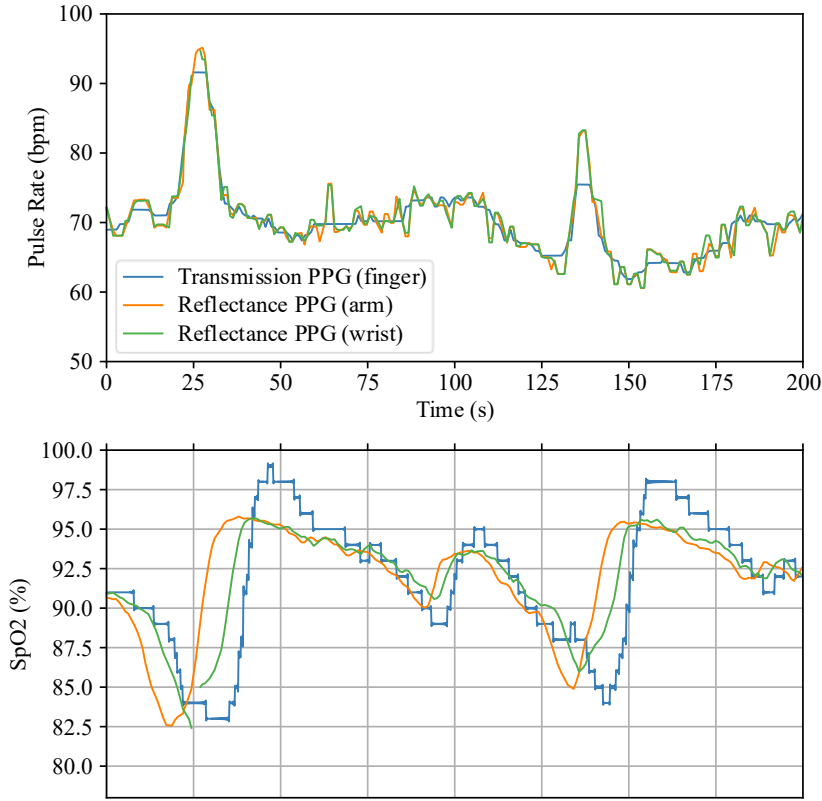

Figure 4 (Top) Offline pulse rate estimate from (blue) ECG reference device, (orange) upper arm wearable and (green) wrist wearable. (Bottom) $SpO_2$ estimate from (blue) fingertip PPG reference, (orange) upper arm wearable and (green) wrist wearable. Both plots use vitals derived from signals synchronized according to inter-beat intervals as described in Section II.B, without accounting for delays in $SpO_2$ estimates. It is clear that the pulse rate estimates match quite well across the devices but that $SpO_2$ outputs exhibit a delay which is location-dependent despite the good signal synchronization.

## C. Different Time Delays of $SpO_2$ at Wrist vs Upper Arm

Another noticeable difference between $SpO_2$ estimation at the upper arm vs. the wrist concerns the reaction time to changes in $SpO_2$ levels. As mentioned in the results section, the delay between the fingertip $SpO_2$ and wearable estimates is 9 seconds for the upper arm and 5 seconds for the wrist. Thus, the upper arm $SpO_2$ estimate seems to react 4 seconds ahead of the wrist $SpO_2$ estimate, underlining an advantage in earlier $SpO_2$-drop detection at the upper arm compared to the wrist. To illustrate this, we provide an example in Figure 4. We hypothesize this to be related to the difference in distance between each of the two measurement locations and the heart, as already observed in previous studies where $SpO_2$ measured at ear, finger and foot showed different delays [8], [9].

## D. Limitations and Future Work

In the current paper, we restrict our analysis to the performance of $SpO_2$ estimation. However, to diagnose SA, these estimates are usually further processed to obtain scores assessing amplitude and duration of $SpO_2$ fluctuations (e.g., oxygen desaturation index [ODI])[5]. Future work should investigate whether wearable-derived $SpO_2$ estimates are sufficient to compute such scores with sufficient reliability when compared to fingertip reference $SpO_2$. In addition, $SpO_2$ estimates should be combined with other PPG-derived vital signs, to provide an unobtrusive solution for long-term monitoring of SA, as well as other diseases involving respiratory symptoms.

## VI. Conclusion

We have evaluated the performance of wearable devices for continuous $SpO_2$ estimation in 134 patients with suspected SA using reflectance PPG data collected at the upper arm and wrist. When compared with the fingertip reference, wearable-derived $SpO_2$ estimations are compliant with both the ISO standard and FDA guidance: $A_{RMS}$ = 3.2% at wrist vs. $A_{RMS}$ = 1.9% at upper arm. However, when compared to the upper arm, we observe a higher rejection rate when estimating $SpO_2$ at the wrist. We attribute this finding to a lower signal quality due to a looser fit at the wrist as well as more frequent and stronger influences from movement. Future work should investigate whether the resulting $SpO_2$ estimates at each measurement location are of sufficient quality to detect and classify SA.

## Acknowledgements

This work was supported by the Swiss National Science Foundation (SNF) project ApneaSense (205321_219441).